\documentclass[conference]{IEEEtran}
\IEEEoverridecommandlockouts
\usepackage{textcomp}
\usepackage{xcolor}
\usepackage{cite}
\usepackage{latexsym}
\usepackage{graphicx}
\usepackage{amsfonts,amssymb,amsmath}
\usepackage{nccmath}
\usepackage{optidef}
\usepackage{hyperref}
\hypersetup{nolinks=true}
\usepackage{url}
\usepackage{color,soul}
\usepackage{array}
\usepackage{algorithm}
\usepackage{algorithmic}
\usepackage[T1]{fontenc}
\usepackage[utf8]{inputenc}
\usepackage{fancyhdr}
\usepackage{lastpage}
\usepackage{caption}
\usepackage{subcaption}

\DeclareMathAlphabet{\pazocal}{OMS}{zplm}{m}{n}

\newcommand{\Mb}{\pazocal{M}}

\newcommand{\Bb}{\pazocal{B}}

\newcommand{\Ib}{\pazocal{I}}

\def\BibTeX{{\rm B\kern-.05em{\sc i\kern-.025em b}\kern-.08em
    T\kern-.1667em\lower.7ex\hbox{E}\kern-.125emX}}
\begin{document}
\title{Impact of NOMA and CoMP Implementation Order on the Performance of Ultra-Dense Networks
}

\author{
    \IEEEauthorblockN{Akhileswar Chowdary \IEEEauthorrefmark{1}, Garima Chopra \IEEEauthorrefmark{1}, Abhinav Kumar\IEEEauthorrefmark{1}, and Linga Reddy Cenkeramaddi\IEEEauthorrefmark{2}}
    \IEEEauthorblockA{\IEEEauthorrefmark{1} Department of Electrical Engineering, Indian Institute of Technology Hyderabad, Telangana, 502285 India.
    \\Email: ee19mtech11028@iith.ac.in, \{garima.chopra, abhinavkumar\}@ee.iith.ac.in}
    \IEEEauthorblockA{\IEEEauthorrefmark{2}Department of Information and Communication Technology, University of Agder, Grimstad, 4879 Norway.
    \\Email: linga.cenkeramaddi@uia.no}
}
\maketitle
\begin{abstract}
Non-orthogonal multiple access (NOMA) is a promising multiple access technology to improve the throughput and spectral efficiency of the users for 5G and beyond cellular networks. Similarly, coordinated multi-point transmission and reception (CoMP) is an existing technology to improve the coverage of cell-edge users. Hence, NOMA along with CoMP can potentially enhance the throughput and coverage of the users. However, the order of implementation of CoMP and NOMA can have a significant impact on the system performance of Ultra-dense networks (UDNs). Motivated by this, we study the performance of the CoMP and NOMA based UDN by proposing two kinds of user grouping and pairing schemes that differ in the order in which CoMP and NOMA are performed for a group of users. Detailed simulation results are presented comparing the proposed schemes with the state-of-the-art systems with varying user and base station densities. Through numerical results, we show that the proposed schemes can be used to achieve a suitable coverage-throughout trade-off in UDNs.
\end{abstract}


\section{Introduction}
Coordinated multi-point transmission and reception (CoMP) has emerged as one of the key technologies for fifth generation (5G) and beyond 5G communications. In the downlink CoMP system, the base stations (BSs) in a CoMP cluster jointly assign dedicated resources to cell-edge users and prohibits the use of the same resources by other non-CoMP users \cite{yogi}. Thus, the throughput of the network reduces due to joint transmission CoMP \cite{yogi}. However, the loss in throughput due to CoMP can be compensated by considering Non-orthogonal multiple access (NOMA). In NOMA, the two users associated with a BS with suitable difference in channel gains can be paired in the power domain. The superposition coding (SC) and successive interference cancellation (SIC) are used in NOMA at the transmitter and receiver, respectively \cite{mouni}.

In an ultra-dense network (UDN), large number of small cells are deployed which in turn reduces the proximity between users and BSs. This BS densification improves the overall spectral efficiency at the cost of increased interference from neighboring BSs \cite{l20}. There are several existing works which have considered NOMA \cite{l3,l5,l15} and CoMP \cite{l19}, individually, for UDN. However, the CoMP and NOMA together have not been studied in detail for UDN. There are key implementation issues, such as grouping, user pairing, and the order in which NOMA and CoMP are implemented which are non-obvious. As mentioned in \cite{aaa}, a CoMP user cannot act as both strong and weak users when paired with multiple non-CoMP users. Given such conditions, pairing of CoMP users and non-CoMP users is a non-trivial task. Recently, in \cite{ourpaper}, a user grouping and pairing scheme for a CoMP–NOMA-based system has been considered for typical user and BS densities. However, the performance also depends on the order of implementation of CoMP and NOMA (NOMA--CoMP, CoMP--NOMA, etc.). Motivated by this, we investigate the effect of CoMP and NOMA implementation order on the UDN by proposing multiple user grouping and pairing schemes. The main contributions of this paper are as follows:
\begin{enumerate}
    \item We propose user grouping and pairing schemes that differ in the order of implementation of NOMA and CoMP and the types of permissible user pairs.
    \item We analyze the CoMP and NOMA based systems using a proportionally-fair scheduler. We believe this is the first paper that considers the proportionally fair scheduling for a CoMP and NOMA based UDN.
    \item We investigate the effect of average cluster size and CoMP signal-to-interference-plus-noise ratio (SINR) threshold on the average throughput of proposed system.
    \item We present detailed simulation results comparing the performance of the proposed schemes with the state-of-art schemes for various user and BS densities.
\end{enumerate}
The organization of the paper is as follows. Section \ref{system_model} describes the system model in detail. Section \ref{pairing} presents the user grouping and pairing schemes proposed in this paper. The simulation results are discussed in Section \ref{results}. The paper is concluded in Section \ref{conclusion}.

\section{System Model} \label{system_model}
\begin{figure}[t]
    \centering
    \includegraphics[width=8.5cm,height=10cm,keepaspectratio]{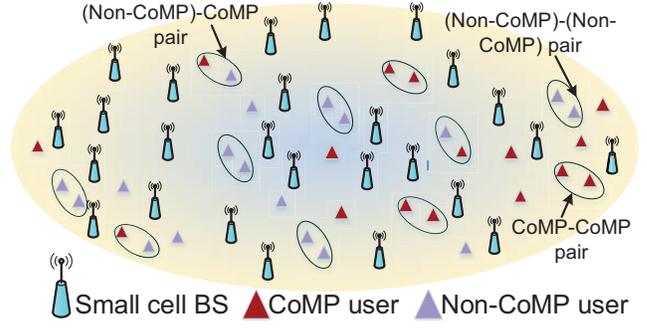}
    \caption{System Model}
    \label{fig:system_model}
\end{figure}

Consider a UDN with users and BSs deployed randomly with densities $\lambda_u$ and $\lambda_b$ following homogeneous Poisson point process (PPP) \cite{ppp} as shown in Fig. \ref{fig:system_model}. Let $\mathcal{M}=\lbrace 1,2,...,M \rbrace$ and $\mathcal{B}=\lbrace 1,2,...,B \rbrace$ be the set of subchannels and BSs, respectively. The users are associated with a BS $b$ based on the maximum received power rule \cite{yogi}.
\subsection{Channel Model}
Assuming Time division duplexing, the downlink SINR of user $i$ on subchannel $m$ from BS $b$ for a maximum transmit power $P^b$ is given as
\begin{equation} \label{eq1}
    \gamma_i^{b,m}=\frac{P^{b,m} g_i^{b,m}}{\sum\limits_{\substack {\hat{b} \neq b \\ \hat{b} \in \Bb}}P^{\hat{b},m}g_{i}^{\hat{b},m} + \sigma^2 } \, ,
\end{equation}
where $P^{b,m} = P^{b}/M$ is the power transmitted per subchannel $m$, $\forall \ m \in \Mb$, $M$ is the total number of subchannels, $g_i^{b,m}$ is the channel gain between user $i$ and BS $b$, ${\sum\limits_{\substack {\hat{b} \neq b \\ \hat{b} \in \Bb}}P^{\hat{b},m}g_{i}^{\hat{b},m}}$ is the interference on subchannel $m$ from neighbouring BSs, and $\sigma^2$ is the noise power. For distance $d_i^b$ between user $i$ and BS $b$, the channel gain can be represented as
\begin{equation} \label{eq2}
    g_i^{b,m}=10^{\frac{-pl(d_i^b)+g_t+g_r-f_s-v}{10}},
\end{equation}
where $pl(d_i^b)$, $g_t$, $g_r$, $f_s$, and $v$ are the path loss of user $i$ at a distance $d^b_i$, transmitter gain, receiver gain, shadowing loss, and penetration loss, respectively. The link rate of a user $i$ with respect to BS $b$ is given as
\begin{equation} \label{eq3}
   r_i^{b}=\frac{\eta(\gamma_i^{b,m}) sc_o sy_o}{t_{sc}} M,
\end{equation}
where $\eta(\gamma_i^{b,m})$ is the spectral efficiency of user $i$ from Adaptive Modulation and Coding Scheme as in \cite{yogi}. Further, $sc_o$, $sy_o$, and $t_{sc}$ represent the number of subcarriers per subchannel, the number of symbols per subchannel, and subframe duration (in seconds), respectively.

\subsection{CoMP}
We consider $\mathcal{C}=\lbrace 1,2,...,C \rbrace$ as the set of CoMP clusters in the area under consideration. We consider \textit{K-means} approach for cluster formation \cite{kmeans}. However, any other clustering can also be used. Let the set of BSs in the CoMP cluster $c$ be denoted by $\mathcal{B}_c$, $\mathcal{B}_c=\lbrace 1,2,...,B_c \rbrace$. For a cluster $c$, the CoMP and non-CoMP users are decided based on the SINR threshold ($\gamma_{th}$). If $\gamma_i^{b,m}<\gamma_{th}$, then user $i$ is designated as CoMP user, otherwise it is treated as a non-CoMP user. The SINR of a CoMP user $i$ in a cluster $c$ is given by \cite{yogi}
\begin{equation} \label{eq4}
    \gamma_{i}^{c,m}=\dfrac{\sum\limits_{\substack{l \in \Bb_{c}}}P^{l,m}g_{i}^{l,m}}{\sum\limits_{\substack{\hat{l} \in \Bb \\ \hat{l} \not\in \Bb_{c}}}P^{\hat{l},m}g_{i}^{\hat{l},m} + \sigma^{2}}, \,
\end{equation}
where ${\sum\limits_{\substack{\hat{l} \in \Bb \\ \hat{l} \not\in \Bb_{c}}}P^{\hat{l},m}g_{i}^{\hat{l},m}}$ is the interference from BSs of neighbouring clusters, $\sum\limits_{\substack{l \in \Bb_{c}}}P^{l,m}g_{i,c}^{l,m}$ is the received power of user $i$ from all BSs in cluster $c$. Let $\theta_c$ be the time duration for which CoMP user $i$ jointly receives information from all BSs in cluster $c$ \cite{yogi}, then the resultant downlink rate, denoted by $\lambda_i^{c}$, is given by
\begin{equation}\label{eq6}
    \lambda_{i}^{c} = \theta_{c}\beta_{i}^{c}r_{i}, \forall i \in \mathcal{I}_c,\,
\end{equation}
where $\mathcal{I}_c$ is the set of CoMP users in cluster $c$, $\beta_{i}^{c}$ is the downlink time fraction for which scheduler assigns all $M$ subchannels to user $i$, and $r_{i}^{c}$ is the link rate of CoMP user $i$. Similarly, the downlink rate for a non-CoMP user is
\begin{equation}\label{61}
    \lambda_{i}^{b} = (1-\theta_{c})\beta_{i}^{b}r_{i}^{b}, \forall i \in \mathcal{I}_{nc}\;\text{and}\;\forall b \in \mathcal{B}_{c},\,
\end{equation}
where $\mathcal{I}_{nc}$ is the set of non-CoMP users in a cluster, $\beta_{i}^{b}$ is the user scheduling time fraction for BS $b$, and $r_{i}^{b}$ is as in \ref{eq3}. The optimal user scheduling time fraction for CoMP and non-CoMP users, $\beta_i^{c}$ and $\beta_{i}^{b}$, respectively, are computed as in \cite{yogi}.

\subsection{NOMA}
We consider power-domain NOMA along with CoMP for an ultra-dense network. The two users are paired as per NOMA scheme based on the minimum SINR criterion as in \cite{mouni}.
Let $\gamma_w^{b,m}$ and $\gamma_s^{b,m}$ be the OMA SINR of weak user $w$ and strong user $s$ computed using (\ref{eq1}), respectively. Then,
\begin{equation} \label{eq7a}
    \hat{\gamma}_s^{b,m}=\frac{\zeta_s P^{b,m} g_s^{b,m}}{\sum\limits_{\substack{\hat{b} \in \Bb \backslash b}}P^{\hat{b},m}g_s^{\hat{b},m} + \sigma^{2}},
\end{equation}
\begin{equation} \label{eq7b}
    \hat{\gamma}_w^{b,m}=\frac{(1-\zeta_s) P^{b,m} g_w^{b,m}}{\zeta_s P^{b,m}g_w^{b,m}+\sum\limits_{\substack{\hat{b} \in \Bb \backslash b}}P^{\hat{b},m}g_w^{\hat{b},m} + \sigma^{2}},
\end{equation}
where $\hat{\gamma}_s^{b,m}$ and $\hat{\gamma}_w^{b,m}$ are the SINR of the strong user with perfect SIC and weak user, respectively, after NOMA pairing, $\zeta_s$ is the power fraction allocated to the strong user which is computed as in \cite{mouni}, $P^{b,m}$ is the total power assigned to the NOMA pair, $g_s^{b,m}$ and $g_w^{b,m}$ is the channel gain of strong and weak user, respectively. In this paper, we have considered Adaptive User Pairing algorithm (AUP) proposed in \cite{mouni}. Next, we explain the proposed NOMA and CoMP schemes.

\section{NOMA and CoMP for UDN} \label{pairing}
There are three kinds of NOMA pairs possible based on the users present in a cluster: CoMP--CoMP, (non-CoMP)--CoMP, and (non-CoMP)--(non-CoMP). We propose two pairing schemes to analyze the performance of CoMP and NOMA based UDN. We also use the scheme proposed in \cite{ourpaper} to study the CoMP and NOMA based UDN.

\subsection{\textit{Scheme A}}
While NOMA increases the throughput of the system, CoMP increases SINR/throughput for cell-edge users \cite{yogi}. Motivated by this, in this scheme, we implement NOMA first $\forall b \in \Bb_c$ to enhance the throughput of the system and then implement CoMP for the unpaired users to enhance their SINR. We pair the users in cluster $c$ using AUP as given in \cite{mouni}. After the implementation of NOMA, we consider all the unpaired OMA users (if any) as CoMP users, irrespective of their SINRs (\textit{we do not follow the CoMP SINR threshold criteria to designate CoMP users in this Scheme}). We then pair the CoMP users using the same AUP algorithm resulting in the formation of CoMP--CoMP NOMA pairs and CoMP OMA users (if any). Thus, in this particular scheme, we have (non-CoMP)--(non-CoMP) NOMA pairs, CoMP--CoMP NOMA pairs, and CoMP OMA users (if any). As all unpaired users are considered as CoMP users, no non-CoMP OMA user exists in this scheme.

The SINRs of a strong and weak user in (non-CoMP)--(non-CoMP) NOMA pair are computed using (\ref{eq7a}) and (\ref{eq7b}), respectively. Similarly, the SINR expressions for the strong user with perfect SIC ($\Bar{\gamma}_{i_s}^{c,m}$) and weak user ($\Bar{\gamma}_{i_w}^{c,m}$) in a CoMP--CoMP NOMA pair are, respectively, given as
\begin{equation} \label{c1}
    \Bar{\gamma}_{s}^{c,m}= \frac{\sum\limits_{\substack{t \in \Bb_{c}}}\zeta_t P^{b,m} g_{s}^{t,m}}{\sum\limits_{\substack{\hat{l} \in \Bb \\ \hat{l} \not\in \Bb_{c}}}P^{\hat{l},m}g_{i_s}^{\hat{l},m} + \sigma^{2}},
\end{equation}
\begin{equation} \label{c2}
    \Bar{\gamma}_{w}^{c,m}= \frac{\sum\limits_{\substack{t \in \Bb_{c}}} (1-\zeta_t) P^{b,m} g_{w}^{t,m}}{\sum\limits_{\substack{t \in \Bb_{c}}} \zeta_t P^{b,m}g_{w}^{t,m} + \sum\limits_{\substack{\hat{l} \in \Bb \\ \hat{l} \not\in \Bb_{c}}}P^{\hat{l},m}g_{i_w}^{\hat{l},m} + \sigma^{2}},
\end{equation}
where $\zeta_t$ is the fraction of the power assigned to strong user in a CoMP--CoMP NOMA pair, $\sum\limits_{\substack{t \in \Bb_{c}}} \zeta_t P^{b,m}g_{w}^{t,m}$ is the interference due to the strong user. The SINR of the CoMP OMA user is computed using (\ref{eq4}). The CoMP--CoMP NOMA pairs and CoMP OMA users are scheduled in the time fraction $\Bar{\theta}_c$, whereas, the (non-CoMP)--(non-CoMP) NOMA pairs of each BS are scheduled in the time fraction of ($1-\Bar{\theta}_c$). The users are scheduled in their respective time fractions using a proportionally-fair scheduler \cite{yogi}. Let $\Bar{\beta}_i^{c}$ be the scheduling time fraction for CoMP--CoMP NOMA pairs and CoMP OMA users and $\Bar{\beta}_{i}^{b}$ be the scheduling time fraction of (non-CoMP)--(non-CoMP) NOMA pairs of BS $b$. We use \cite{yogi} to compute $\Bar{\theta}_c$, $\Bar{\beta}_i^{c}$, and $\Bar{\beta}_{i}^{b}$ that are as follows
\begin{figure}[t]
    \centering
    \includegraphics[width=9cm,height=10.5cm,keepaspectratio]{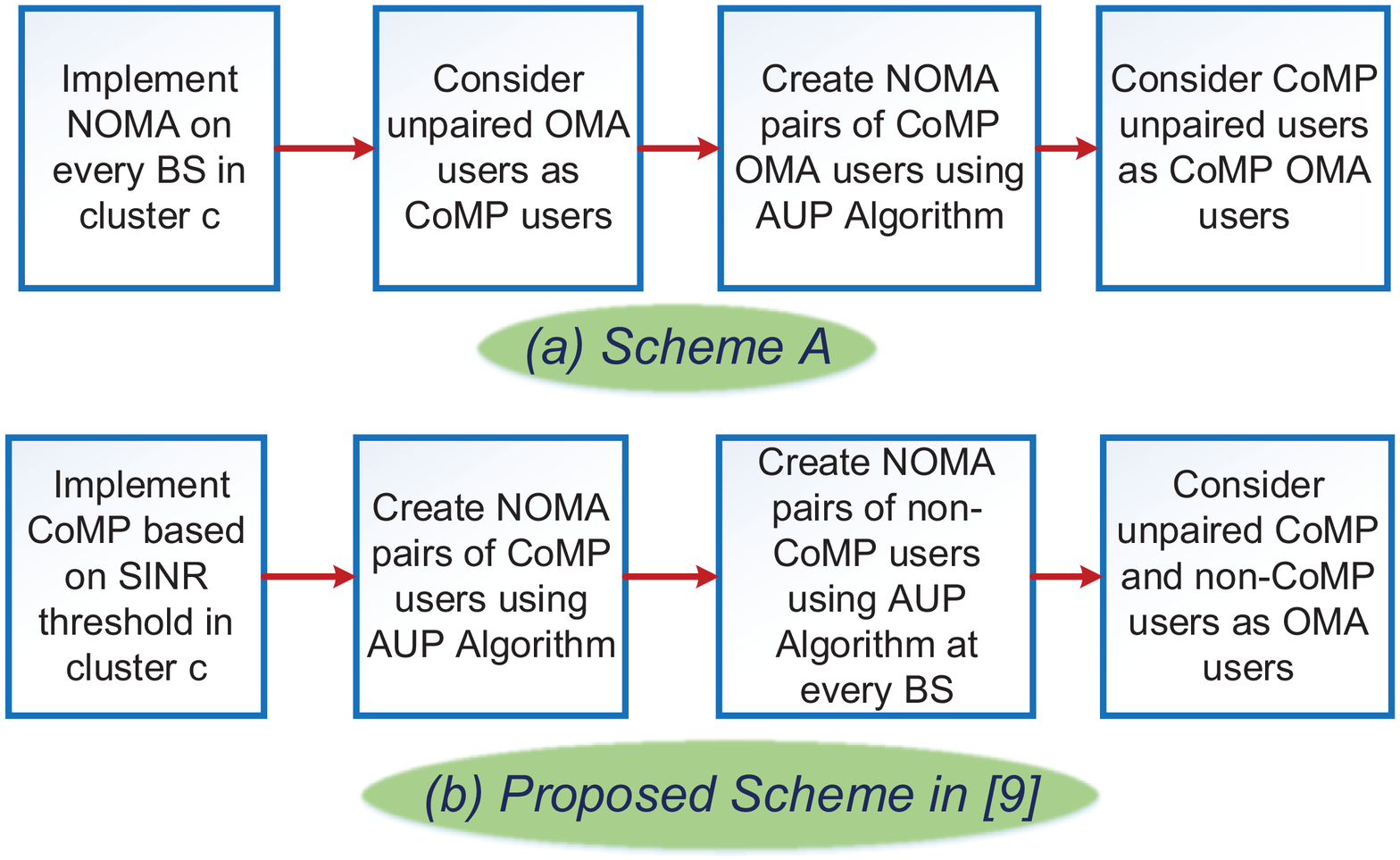}
    \caption{Illustration of \textit{Scheme A}.}\vspace{-0.1in}
    \label{fig:scheme_a}
\end{figure}
\begin{equation}
    \Bar{\theta}_c = \frac{|\Bar{\Ib}_c| + |\Bar{\Ib}_c^{oma}|}{|\Bar{\Ib}_c| + |\Bar{\Ib}_c^{oma}| + |\Bar{\Ib}_{nc}|},\,
\end{equation}
\begin{equation}
    \Bar{\beta}_{i}^{c} = \frac{1}{|\Bar{\Ib}_c| + |\Bar{\Ib}_c^{oma}|},\, \mbox{and} \, \Bar{\beta}_{i}^{b} = \frac{1}{|\Bar{\Ib}_{nc}^{b}|},\,
\end{equation}
where $|X|$ represents the cardinality of set $X$, $\Bar{\Ib}_c$, $\Bar{\Ib}_c^{oma}$, and $\Bar{\Ib}_{nc}$ are the set of CoMP NOMA pairs, CoMP OMA (unpaired) users, and non-CoMP NOMA pairs, respectively, in cluster $c$, and $\Bar{\Ib}_{nc}^{b}$ is the set of (non-CoMP)--(non-CoMP) pairs formed with the users associated with the BS $b$. These results are derived in \cite{yogi} for a purely CoMP system, hence, they may be sub-optimal for this scheme. A detailed schematic of \textit{Scheme A} is presented in Fig. \ref{fig:scheme_a}(a).

\subsection{\textit{Scheme B}}

In this scheme, we implement CoMP first for a cluster $c$ to get the cell edge users under coverage and then implement NOMA to boost their rates. To avoid complexities while pairing and to ensure that a CoMP user paired with multiple users acts as either strong or weak user with all the users \cite{aaa}, in this scheme, we consider (non-CoMP)--CoMP pairs such that CoMP user is always a weak user in the pair formed. This also abstains the CoMP user from performing SIC. Similar pairing can also be performed with CoMP user as strong user at the cost of increased receiver complexity. Therefore, in this scheme, we consider only (non-CoMP)--CoMP with CoMP user as weak user and (non-CoMP)--(non-CoMP) NOMA pairs to study the CoMP and NOMA based UDN. Firstly, all the users in cluster $c$ are divided into groups $\textbf{G1}$ and $\textbf{G2}_b$, where $\textbf{G1}$ contains the SINRs of CoMP users in cluster $c$, $\textbf{G1}=\lbrace \gamma_{1,c}^m, \gamma_{2,c}^m,...,\gamma_{i,c}^m \rbrace$ and $\textbf{G2}_b$, $\textbf{G2}_b=\lbrace \gamma_1^{b,m}, \gamma_2^{b,m},...,\gamma_i^{b,m} \rbrace$ contains the SINRs of non-CoMP users associated with BS $b$ in the cluster $c$. $\textbf{G2}_b$ is formed for every BS $b$ in cluster $c$. To apply AUP algorithm, we need to form two user groups, namely, a weak user group and a strong user group. The SINR of every user in the weak user group should be less than that of every user in the strong user group. Therefore, in this scheme, the necessary condition for pairing the CoMP users in $\textbf{G1}$ is that there should exist atleast one user in the \textbf{G1} whose SINR is less than the maximum of SINRs of all users in $\textbf{G2}_b$. The CoMP users which satisfy this condition are eligible to be paired with users in $\textbf{G2}_b$ for a BS $b$. After verifying this condition, we first form a new group $\textbf{G1}_b^{'}$ with those users in \textbf{G1} that satisfy the previously mentioned condition. After forming $\textbf{G1}_b^{'}$, the group $\textbf{G2}_b^{'}$ is formed for each BS $b$ by picking those users from $\textbf{G2}_b$ whose SINR is greater than maximum SINR of all users in $\textbf{G1}_b^{'}$. The aforementioned procedure is carried out for every BS present in the cluster in an iterative manner till there exists no CoMP user whose SINR is less than atleast one non-CoMP user in every $\textbf{G2}_b$ formed. Then, the users in both the groups are paired using AUP. Therefore, a CoMP user can be a part of NOMA pairs of multiple BSs simultaneously. The unpaired CoMP users are served as OMA users. The non-CoMP users associated with a BS $b$ that are not paired with the CoMP users are paired among themselves using the same AUP at every BS. The SINR of the weak CoMP user ($\Tilde{\gamma}_{w}^{c,m}$) in a (non-CoMP)--CoMP NOMA pair is computed using (\ref{eq4}) and (\ref{eq7b}) and given as follows.
\begin{equation} \label{eq10}
\Tilde{\gamma}_{w}^{c,m}= \frac{\sum\limits_{\substack{k \in \Tilde{\Bb}_c}} (1-\zeta_k) P^{b,m} g_{w}^{k,m} + \sum\limits_{\substack{q \in  \Bb/\Tilde{\Bb}_c}} P^{b,m} g_{w}^{q,m}}{\sum\limits_{\substack{k \in \Tilde{\Bb}_c}} \zeta_k P^{b,m}g_{w}^{k,m} + \sum\limits_{\substack{\hat{l} \in \Bb \\ \hat{l} \not\in \Bb_{c}}}P^{\hat{l},m}g_{w}^{\hat{l},m} + \sigma^{2}},
\end{equation}
where $\Tilde{\Bb}_c$ is the subset of BSs in $\Bb_c$ with which a CoMP user has formed pairs. Similarly, the SINR of the strong non-CoMP user ($\Tilde{\gamma}_{s}^{k,m}$) in a (non-CoMP)--CoMP NOMA pair with perfect SIC is given as follows.
\begin{equation} \label{eq11}
    \Tilde{\gamma}_{s}^{k,m}= \frac{\zeta_k P^{k,m} g_{s}^{k,m}}{\sum\limits_{\substack{\hat{k} \in \Bb \backslash k}}P^{\hat{k},m}g_{s}^{\hat{k},m} + \sigma^{2}}, \forall k \in \Bb_c, 
\end{equation}
where $\zeta_k$ is the power fraction allocated by the BS $k$ in the cluster $c$. In addition, one more kind of pairing considered in this scheme is (non-CoMP)--(non-CoMP). The SINR of users involved in (non-CoMP)--(non-CoMP) pairing can be computed using (\ref{eq7a}) and (\ref{eq7b}).
\begin{figure}[t]
     \centering
     \includegraphics[width=9cm,height=11cm,keepaspectratio]{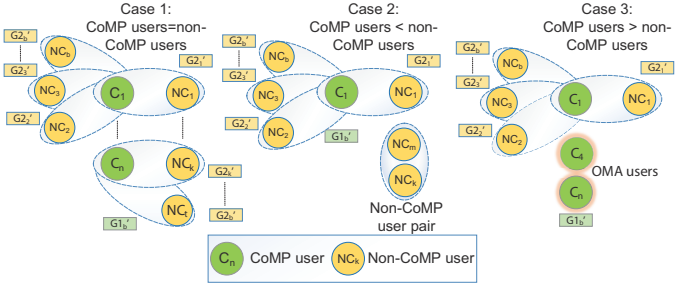}
     \caption{User grouping and pairing based on \textit{Scheme B}.}\vspace{-0.19in}
     \label{fig:scheme_1}
\end{figure}

Note that a CoMP user can be paired with more than one non-CoMP user (each non-CoMP user associated with different BS in cluster $c$) and thus, forming a NOMA-CoMP cluster as shown in Fig. \ref{fig:scheme_1}. (non-CoMP)--CoMP NOMA pairs and CoMP OMA users (if any) are scheduled in the CoMP time fraction of $\Tilde{\theta}_c$. Each pair is scheduled using a proportionally-fair scheduler. For illustration, suppose that a CoMP user $i \in \Ib_{c}$ is paired with 3 non-CoMP users associated with 3 different BSs in the cluster $c$. The CoMP user $i$ are served by all the BSs in the cluster in the duration of $\Tilde{\theta}_c$ within the scheduled time fraction of $\Tilde{\beta}_{i}^{c}$. In the same duration, the 3 non-CoMP users are served by their respective BSs. During the remaining $(1-\Tilde{\theta_c})$ duration, each BS in cluster $c$ schedules their respective non-CoMP user pairs and non-CoMP OMA users (if any) using a proportionally-fair scheduler within the scheduled time fractions \{$\Tilde{\beta}_{i}^{b}$\}. The non-CoMP users that are paired with CoMP users are not be served in the duration of $(1-\Tilde{\theta_c})$. The $\Tilde{\theta}_c$, $\Tilde{\beta}_{i}^{c}$, and $\Tilde{\beta}_{i}^{b}$ using \cite{yogi} are given, respectively, as follows.
\begin{equation}\label{eq:comp_tf1}
    \Tilde{\theta}_c = \frac{|\Tilde{\Ib}_c| + |\Tilde{\Ib}_{c}^{oma}|}{|\Tilde{\Ib}_c| + |\Tilde{\Ib}_{c}^{oma}| + |\Tilde{\Ib}_{nc}| + |\Tilde{\Ib}_{nc}^{oma}| },
\end{equation}
\begin{equation}\label{eq:beta_1}
    \Tilde{\beta}_{i}^{c} = \frac{1}{|\Tilde{\Ib}_c| + |\Tilde{\Ib}_{c}^{oma}|},\, \mbox{and}\, \Tilde{\beta}_{i}^{b} = \frac{1}{|\Tilde{\Ib}_{nc}^{b}| + |\Tilde{\Ib}_{nc}^{b,oma}|},\,
\end{equation}
where $\Tilde{\Ib}_c$ is the set of (non-CoMP)--CoMP users in cluster $c$, $\Tilde{\Ib}_{c}^{oma}$ is the set of OMA CoMP users in cluster $c$, $\Tilde{\Ib}_{nc}$ is the set of (non-CoMP)--(non-CoMP) pairs in cluster $c$, $\Tilde{\Ib}_{nc}^{oma}$ is the set of OMA non-CoMP users in the cluster $c$, $\Tilde{\Ib}_{nc}^{b}$ is the set of (non-CoMP)--(non-CoMP) pairs associated with the BS $b$, and $\Tilde{\Ib}_{nc}^{b,oma}$ is the set of OMA non-CoMP users associated with the BS $b$.

\subsection{\textit{Scheme Proposed in \cite{ourpaper}}}
The scheme in \cite{ourpaper} differs from \textit{Scheme B} in the types of user pairs formed. Further the scheme in \cite{ourpaper} has been proposed for a typical cellular scenario, whereas, we evaluate this scheme's performance for a UDN in this paper. In this scheme only CoMP--CoMP and (non-CoMP)--(non-CoMP) NOMA pairs are formed. The SINRs of weak and strong CoMP user in the CoMP--CoMP NOMA pair are given in \cite{ourpaper}. Similarly, The SINRs of strong and weak users of (non-CoMP)--(non-CoMP) NOMA pair can be computed using (\ref{eq7a}) and (\ref{eq7b}), respectively. In this scheme, there is no scope for a NOMA-CoMP cluster as in \textit{Scheme B}.
The CoMP--CoMP pairs are scheduled in the duration of $\Hat{\theta}_c$. Each CoMP--CoMP NOMA pair or OMA CoMP user (if any) is given a time fraction of $\Hat{\beta}_{i}^{c}$, whereas (non-CoMP)--(non-CoMP) NOMA pairs are served by their respective BSs in the duration of $(1-\Hat{\theta}_c)$ with a proportionally fair scheduler and each pair is given a time fraction of $\Hat{\beta}_{i}^{b}$. The expressions of $\Hat{\theta}_c$, $\Hat{\beta}_{i}^{c}$, and $\Hat{\beta}_{i}^{b}$ are given in \cite{ourpaper}. A detailed schematic of this scheme is presented in Fig. \ref{fig:scheme_a}(b).
\section{Results and Discussions} \label{results}
\begin{table}[t]
\caption{Simulation Setup}
\begin{center}
\begin{tabular}{| m{4.4cm} | m{3.3cm}|}
\hline
\textbf{\textit{Parameter}}& \textbf{\textit{Value}} \\
\hline
Area under consideration ($\text{km}^2$) & 1\\
\hline
AWGN Power spectral density (dBm) & $-174$ \\
\hline
Base Station density, $\lambda_{b}$ (/$\text{km}^2$) & $100,200,300,400,500$\\
\hline
Number of subcarriers per subchannel, $sc_o$ & $12$ \\
\hline
Number of symbols per subcarrier, $sy_o$ & $14$ \\
\hline
Average Cluster Size & $5,8,10$ \\
\hline
Number of iterations & $10^5$ \\
\hline
Standard deviation of shadowing random variable (dB) & 8\\
\hline
Subchannel Bandwidth (kHz) & $180$ \\
\hline
Total number of subchannels, M & $100$ \\
\hline
Transmission power, $P^{b}$ (dBm) & $24$ dBm \\
\hline
User density, $\lambda_u$ (/$\text{km}^2$) & $100,200,400,600,1000$\\
\hline
\end{tabular}\vspace{-0.15in}
\label{table}
\end{center}
\end{table}

\begin{figure}[t]
     \centering
     \includegraphics[width=9.5cm,height=12cm,keepaspectratio]{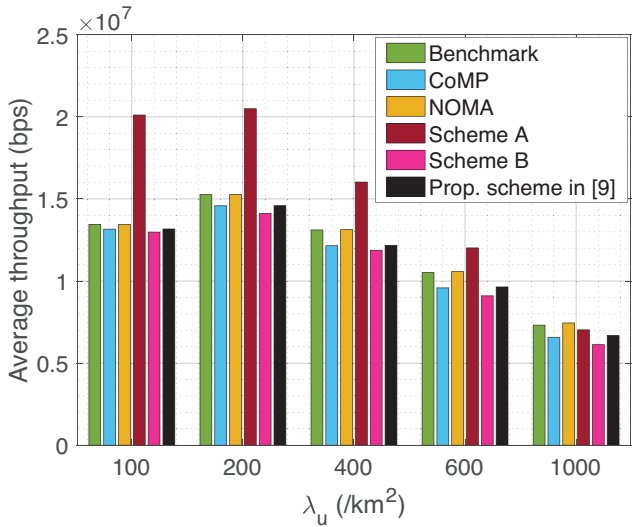}
     \vspace{-0.15in}
     \caption{Variation of throughput with $\lambda_u$ for $\lambda_{b} = 500/\text{km}^2$.}
     \label{fig:throughput_1} \vspace{-0.25in}
\end{figure}

The parameters considered for simulation are summarized in Table \ref{table}. In this paper, we consider the expressions of average throughput and coverage as in \cite{yogi}. Fig. \ref{fig:throughput_1} shows the variation of average throughput with respect to $\lambda_u$ for $\lambda_b = 500/\text{km}^2$. It is observed that for $\lambda_u \leq 600$/km$^2$, \textit{Scheme A} outperforms \textit{Scheme B} and the scheme proposed in \cite{ourpaper} as well as conventional schemes. This is due to the formation of a very few NOMA pairs (or even no NOMA pairs) at the initial stage of NOMA implementation, particularly for lower $\lambda_u$. As we do not use $\gamma_{th}$ to separate CoMP users in \textit{Scheme A}, a large number of unpaired OMA users are considered as CoMP users unlike in \textit{Scheme B} and the scheme in \cite{ourpaper}. The pairing of such CoMP users using NOMA results in the increase in the average throughput of \textit{Scheme A}. For a given $\lambda_b$, at relatively higher $\lambda_u$, number of users per BS increases which leads to an increase in the number of non-CoMP NOMA pairs and decrease in the unpaired non-CoMP OMA users. Therefore, at higher $\lambda_u$, NOMA is performing better than \textit{Scheme A.}

\begin{figure}[t]
     \centering
     \includegraphics[width=8.5cm,height=10.5cm,keepaspectratio]{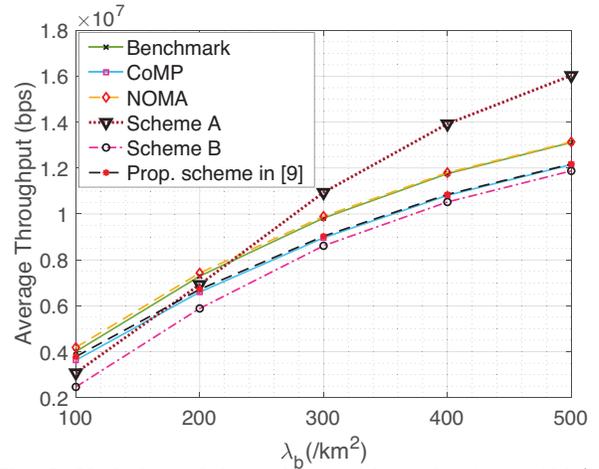}
     \vspace{-0.15in}
     \caption{Variation of throughput with $\lambda_b$ for $\lambda_{u} = 400/\text{km}^2$.}
     \label{fig:throughput_2} \vspace{-0.1in}
\end{figure}

We present the variation of the average throughput of the system for various $\lambda_b$ and for $\lambda_u = 400$/km$^2$ in Fig. \ref{fig:throughput_2}. For a given $\lambda_u$, at comparatively lower $\lambda_b$, more number of users can get associated with a single BS. Therefore, there is a higher possibility for non-CoMP user NOMA pairs being formed. However, with increase in $\lambda_b$, number of users per BS decreases. Therefore, the possibility of non-CoMP NOMA pairs being formed decreases. Hence, the number of unpaired non-CoMP OMA users increases with increase in $\lambda_b$. Therefore, the average throughput of \textit{Scheme A} is superior for higher $\lambda_b$ as compared to lower $\lambda_b$. The relatively lower performance of \textit{Scheme B} and the scheme proposed in \cite{ourpaper} in Fig. \ref{fig:throughput_1} and Fig. \ref{fig:throughput_2} can be attributed to the threshold based CoMP implementation in the initial stage. Due to this SINR threshold based CoMP, time fraction available to the users gets reduced because the $\hat{\theta}_c$ and $\Tilde{\theta}_c$ start increasing for lower values of $\lambda_b$. The (non-CoMP)--CoMP pairing in \textit{Scheme B} further reduces the time fraction available for some of the non-CoMP users. Hence, its average throughput performance is the worst when compared to all other schemes. At higher $\lambda_b$, number of users per BS decreases which inturn decreases the number of non-CoMP NOMA pairs. Therefore, for \textit{Scheme B} and the scheme proposed in \cite{ourpaper}, the average throughput is similar to that of CoMP-only system.

\begin{figure}[t]
     \centering
     \includegraphics[width=8.5cm,height=10.5cm,keepaspectratio]{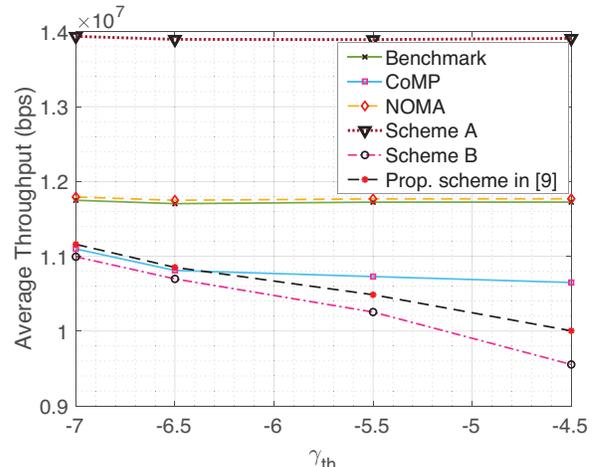}
     \vspace{-0.15in}
     \caption{Variation of throughput with $\gamma_{th}$ for $\lambda_{u} = 200/\text{km}^2$ and $\lambda_{b} = 200/\text{km}^2$.} \vspace{-0.12in}
     \label{fig:throughput_3}
\end{figure}
\begin{figure}[t]
     \centering
     \vspace{-0.11in}
     \includegraphics[width=8.5cm,height=10.5cm,keepaspectratio]{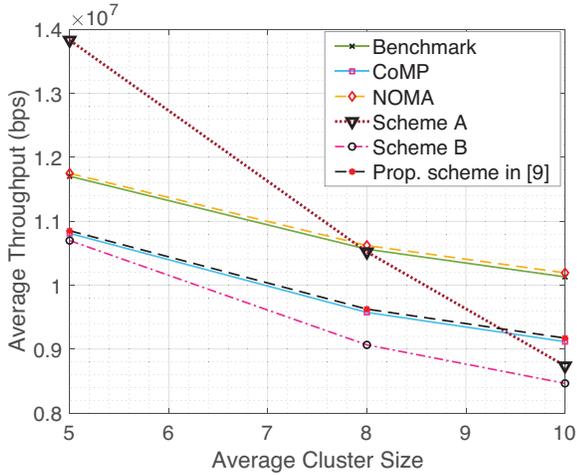}
     \caption{Variation of throughput with the average cluster size for $\lambda_{u} = 200/\text{km}^2$ and $\lambda_{b} = 200/\text{km}^2$.} \vspace{-0.15in}
     \label{fig:throughput_4}
\end{figure}

Fig. \ref{fig:throughput_3} presents the variation of average throughput with respect to $\gamma_{th}$. We have observed that \textit{Scheme A}, benchmark, and NOMA-only systems maintain average throughput as constant for all values of $\gamma_{th}$. Whereas, a downward trend was observed for \textit{Scheme B}, for the scheme proposed in \cite{ourpaper}, and for CoMP only system with an increase in $\gamma_{th}$. However, the drop in the average throughput for CoMP only system is not as steep as it is for \textit{Scheme B} and for the scheme proposed in \cite{ourpaper}. The reason for such marginal drop is that at such high $\lambda_b$, an increase in the CoMP threshold may not increase the number of CoMP users significantly that can make huge differences in throughput. However, there is a marginal increase in the number of CoMP users because of which there is a marginal drop in the average throughput. Nevertheless, for \textit{Scheme B} and for the scheme proposed in \cite{ourpaper}, this marginal increase in the number of CoMP users leads to an increase in the number of (non-CoMP)--CoMP and CoMP--CoMP NOMA pairs, respectively. Therefore, $\Hat{\theta}_c$ and $\Tilde{\theta}_c$ increases due to which the drop in the average throughput is much steeper than that of CoMP.

The variation of average throughput with respect to average CoMP cluster size is shown in Fig. \ref{fig:throughput_4}. As the average cluster size increases, the performance of \textit{Scheme A} starts deteriorating. This is due to the increase in the number of NOMA pairs as well as unpaired OMA users that are considered as CoMP users in \textit{Scheme A}. With the increase in the number of unpaired OMA users, the $\Bar{\theta}_c$ gradually increases. Hence, the performance of \textit{Scheme A} gets worse than that of the scheme proposed in \cite{ourpaper} as the cluster size increases.
\begin{figure}[t]
     \centering
     \vspace{-0.11in}
     \includegraphics[width=8.5cm,height=10.5cm,keepaspectratio]{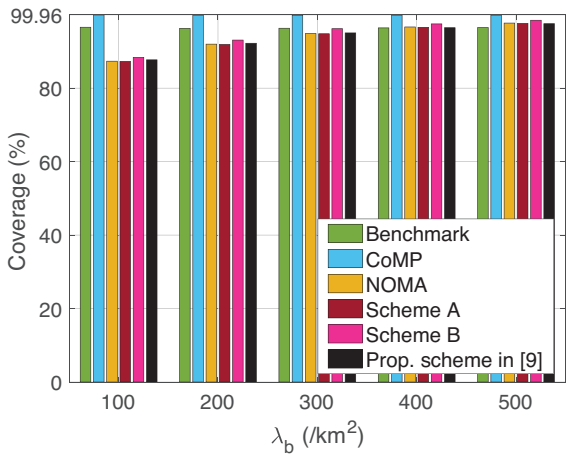}
     \caption{Variation of coverage with $\lambda_b$.}\vspace{-0.15in}
     \label{fig:coverage}
\end{figure}

Fig. \ref{fig:coverage} presents the variation of the coverage of the various schemes with respect to $\lambda_b$. We can observe that the coverage of CoMP and NOMA based systems is less than the Benchmark and CoMP-only systems at lower $\lambda_b$. However, their coverage is comparatively equal or slightly greater than the NOMA-only system. The \textit{Scheme A} is performing better than other schemes in terms of throughput under certain conditions but its coverage is less. The \textit{Scheme B}'s performance in terms of throughput is the worst. However, its coverage is slightly better than NOMA and other two schemes. Thus, the proposed schemes offer a trade-off between the coverage and throughput.

\section{Conclusion} \label{conclusion}
In this paper, we have proposed multiple user grouping and pairing schemes to study the performance of CoMP and NOMA based UDN. The proposed schemes not only differ in the order of implementation of NOMA and CoMP but also in the kinds of permitted NOMA pairs. We have compared the performance of these schemes with the conventional OMA-based benchmark, CoMP-only, NOMA-only systems, and the state-of-the-art. The \textit{Scheme A} among the proposed schemes results in the enhanced throughput when compared to its counterparts for lower $\lambda_u$ and higher $\lambda_b$. The coverage of all the three schemes is less than the benchmark and CoMP-only systems for lower $\lambda_b$. The \textit{Scheme B} performs marginally better than the other schemes and NOMA-only in terms of coverage. The proposed schemes can be used by cellular network planners to appropriately deploy UDNs.

\section{Acknowledgement}
This work was supported in part by the Indo-Norwegian Collaboration in Autonomous Cyber-Physical Systems (INCAPS)--project: 287918 of the INTPART program, the Low-Altitude UAV Communication and Tracking (LUCAT)--project: 280835 of the IKTPLUSS program from the Research Council of Norway and the Department of Science and Technology (DST), Govt. of India (Ref. No. INT/NOR/RCN/ICT/P-01/2018), and DST NMICPS through TiHAN Faculty fellowship of Dr. Abhinav Kumar.

\bibliography{ref_1}
\bibliographystyle{IEEEtran}
\vspace{14mm}

\end{document}